\documentclass[aps,prc,reprint,groupedaddress]{revtex4-1}

\usepackage {graphicx}
\usepackage{amssymb}
\usepackage{amsmath}
\usepackage{units}

\def\mb{\,\mbox{mb}}
\def\fm{\,\mbox{fm}}
\def\GeV{\,\mbox{GeV}}

\def\TeV{\,\mbox{TeV}}

\begin{document}


\title{Universal Centrality Dependence of Particle Multiplicities in Heavy-Ion Collisions}


\author{H.J. Pirner}
\affiliation{Institute for Theoretical Physics, University of Heidelberg, Germany}

\author{K. Reygers}
\affiliation{Physikalisches Institut, University of Heidelberg, Germany}

\author{B.Z. Kopeliovich}
\affiliation{Departamento de F\'{\i}sica
Universidad T\'ecnica Federico Santa Mar\'{\i}a; and
\\
Instituto de Estudios Avanzados en Ciencias e Ingenier\'{\i}a; and\\
Centro Cient\'ifico-Tecnol\'ogico de Valpara\'iso;\\
Casilla 110-V, Valpara\'iso, Chile\\}


\date{\today}

\begin{abstract}
Starting from the light-cone plasma (LCP) distribution of gluons produced in proton-proton collisions, we derive the invariant charged-particle yield $dN/dyd^2p_\perp$ in heavy-ion collisions.  Multiple scattering of partons in the other nucleus leads to $p_{\bot}$ broadening, which depends on the centrality of the collision.  The resulting change of the LCP parameters determines the distributions of charged particles. The charged-particle multiplicity $dN/d\eta$ at $\eta=0$ normalized to the number of participating nucleons has a universal centrality dependence, common to the Relativistic Heavy Ion Collider (RHIC) and the Large Hadron Collider (LHC). This universal increase is of geometrical origin. Such a universality can be broken by the energy-dependent coherence effect, shadowing and mutual boosting of the saturation scales. However, we found these effects to essentially compensate each other, leaving the centrality dependence of multiplicity approximately universal.  
\end{abstract}


\maketitle

\section{Introduction}
\label{sec:intro}

In a recent publication \cite{Pirner:2011ab} a universal multiplicity distribution for produced partons has been deduced from maximum-entropy theory. Given the experimentally measured total transverse energy $E_{\bot}$ and the sum rule requiring the total light-cone fractions $x$ of partons in the proton to be unity, one obtains a Bose-distribution $n(x,p_{\bot})$ for gluons with two Lagrange parameters governing its $x$ and $p_{\bot}$ dependence. The underlying picture is simple. In soft high energy collisions partons interact by exchanging small transverse momenta and become liberated subsequently. The resulting hadronization products reflect the parton distributions via hadron-parton duality. This picture has been worked out for proton-proton (pp) and nucleus-nucleus (AA) collisions in Ref.~\cite {Pirner:2011ab}. It provides a satisfactory explanation for soft particle production in pp collisions throughout the accessible rapidity range. Nuclear multiplicity distributions $dN/dyd^2p_\perp$ can be calculated by scaling the pp distribution with the number of participating nucleons \emph{and} taking into account the higher mean transverse momentum $\langle p_{\bot} \rangle$ observed in experiment. In this paper we want to theoretically calculate this increased mean transverse momentum from parton multiple scattering. 

To describe the inclusive cross section of charged particles produced in pp collisions,
\begin{equation}
\frac{dN}{dy d^2 p_{\bot}}=\frac{dN}{d\ln x d^2 p_{\bot}}
                          =\frac {d \sigma}{ \sigma_{in} dy d^2 p_{\bot}},
\end{equation}
we use a statistical distribution function which maximizes the entropy of the produced partons given certain constraints.  We consider the light-cone momenta of the partons with energies $\epsilon$ and longitudinal momenta $p_{z}$ relative to the light-cone momentum of the incoming proton with $(E,P_{z},0)$:
\begin{equation}
x= \frac{\epsilon+p_{z}}{E+P_{z}} = \frac{p_+}{P_+}.
\end{equation}

The maximum-entropy distribution, i.e., the LCP distribution, has to satisfy the following two requirements: The first requirement concerns light-cone momentum conservation; i.e.\ the sum of all partonic light-cone fractions is unity. The second constraint is given by the measured total transverse energy of the produced partons. For symmetric collisions in the center-of-mass (cm) system, we can consider the multiplicity distribution in each hemisphere separately. The resulting individual distributions function have the form of a Bose-Einstein distribution depending on transverse momentum and light-cone fractional momentum $x$. It is important to emphasize the light-cone property of the maximum-entropy distribution. The dynamics of collisions at high energies is governed by the the light-cone Hamiltonian which is boost-invariant and determines the distribution functions depending on transverse momentum and light-cone fractions which label the eigenstates. The density matrix resulting after the collision can be built up from an incoherent mixture of such multi-parton states. Without a boost invariant formulation one cannot define a number density of partons since in each reference system it will be different. In the rest system of the proton the gluons sit in the clouds around the constituent quarks and the strings holding them together, whereas in a fast-moving proton the virtual gluons materialize as partons carrying a sizable momentum fraction. It is this physics which also determines the maximum-entropy state.

In collisions the projectile and target have to interact, and this interaction takes place between the soft partons with low Bjorken $x$. They acquire transverse momenta; therefore, the produced total transverse energy represents an important input for the maximum-entropy distribution. Boost invariance applies to the change of the measured inclusive distribution in the cm system under a change of cm energy. Once the distribution is formulated with the variables $p_{\bot}$ and $x$ it can be applied to any cm energy. In reality, the mean transverse momentum will increase with increasing cm energy and thereby also modify the softness of the particles in $x$. The maximum-entropy principle can describe a real nonequilibrium state formed in the collision or an average over many collisions.  It is probable that each individual collision has a different transverse momentum spectrum, but on average we observe an effective transverse temperature characterizing the cross section.

The question arises: Is such a formulation really necessary, since we have the very successful Bjorken picture \cite{Bjorken:1982qr} which also includes boost invariance. Indeed in the central region where the rapidity distribution is flat, there is a considerable overlap between the two formulations, so the effective transverse temperature represented by the parameter $\lambda$ in the above distribution is related to the multiplicity distribution for $y=0$ in a similar way. There are also differences. In the light-cone formulation the rapidity distribution is not flat. Furthermore, the applicability of hydrodynamics is a fundamental ingredient of the Bjorken picture. From a very early time $\tau_0\sim0.2\fm$ on, a locally equilibrated fluid has to be assumed which expands longitudinally. More complicated modern hydrodynamical calculations \cite {Kolb:2003dz} allow also transverse expansion with viscosity, and calculate the radial flow parameters which seem to support this picture. A detailed time evolution based on the thermal equation-of-state results. However, it has to be carefully checked whether the many assumptions underlying these calculations are really reflected in the data. Minimal assumptions based on conservation laws, complemented by few key observables may also reproduce the main features of the experiments.

We think that the maximum-entropy distribution can serve such a purpose. Building on the dynamics of the underlying $pp$ events, nuclear collisions represent a superposition of pp events. Multiparton collisions in the evolving larger system increase the partonic transverse momenta which lead to higher hadronic momenta in nuclear collisions. We want to understand the microscopic origin of this transverse broadening and its energy dependence as a function of the centrality of the collision. Thereby we can separate the
geometrical and dynamical aspects of these processes. 

The outline of the paper is as follows. In Sec.~II we review the features of the longitudinal light-cone plasma (LCP) of gluons which we need. In Sec.~III we calculate the mean density profile a ``beam parton'' experiences in a nuclear collision from averaging over ``target nucleons'' and evaluate the resulting $p_{\bot}$ broadening. In Sec.~IV we consider the effects of mutual boosting of saturation scales and shadowing in nuclear collisions and compare with the simple parametrization of Sec.~III. Finally, in Sec.~V the effect of transverse momentum  broadening or saturation on the pseudo-rapidity distributions is calculated using the relevant LCP distributions. The results are discussed in Sec.~VI.

\section{The LCP Distribution from the Maximum-Entropy Principle}

The maximum-entropy principle for the multiplicity distribution in pp collisions can be best derived from the entropy of a Bose system which is not in thermal equilibrium but subject to constraints. The entropy of the system is proportional to the logarithm of the integrated phase space. In the following we set the Boltzmann constant $k=1$ and $\hbar=1$.  On the light cone, phase space includes the transverse spatial coordinate $b_{\bot}$, the transverse momentum $p_{\bot}$, the longitudinal light-cone momentum $p_+$ and the longitudinal spatial variable $z_-=1/2(t-z)$, where $t$ and $z$ are time and longitudinal coordinate.  These variables are handled as in conventional thermodynamics, i.e.\ the phase space multiplied by the gluon degeneracy factor $g=2 (N_c^2-1)$ and divided by the Planck constant $h^3=(2 \pi \hslash)^3$ gives the number of available quantum states $G$:
\begin{eqnarray}
G_{b_{\bot},p_{\bot},p_+,z_-}&=&g  \frac{d^2 b_{\bot} d^2p_{\bot}dp_+dz_-}{(2 \pi)^3} \\
                                  &=&g  \frac{d^2 b_{\bot} d^2p_{\bot}}{(2 \pi)^2} dx \frac{d \rho}{2 \pi}.
\end{eqnarray}

For high energies, Feynman scaling is a good phenomenological concept; therefore, we have multiplied and divided this expression by $P_+=E+P_z$ to obtain the light-cone momentum $x=p_+/P_+$ and the longitudinal light-cone distance $\rho=z_-P_+$ which are canonically conjugate variables.We further make the simplifying assumption that the distribution function and consequently the entropy are homogeneously distributed in transverse space. The integration over the $b_{\bot}$ coordinate can then be executed and gives the area $L_{\bot}^2$.  An estimate of the integral of the scaled light-cone distance $\int \frac{d \rho}{2 \pi}$ is more subtle, since it is not independent on the rest of the variables. It has to be done separately for valence and sea partons.  Interpolating the $x$ dependence of the longitudinal extension of valence and sea partons we obtain the integral over the scaled distance; see Ref.~\cite{Pirner:2011ab}:
\begin{equation}
\int \frac{d \rho}{2 \pi} \approx \frac{1}{x}.
\end{equation}
The so motivated ansatz for the phase space on the light cone is crucial for all further derivations:  
\begin{equation} 
G_{x,p_{\bot}}=g L_{\bot}^2 \frac{ d^2 p_{\bot}}{(2 \pi)^2} \frac{dx}{x}.  
\end{equation}

Gluons are bosons; therefore, they can occupy the phase-space cells in multiples. The binomial of the combined number of particles and states over the number of states gives the number of possibilities $\Delta \Gamma_{x,p_{\bot}}$ to distribute $N_{x,p_{\bot}}$ gluons, i.e.\ bosons, on $G_{x,p_{\bot}}$ quantum states. The entropy of the system is defined by the logarithm of the phase space. For large particle numbers and quantum states one can express the entropy in terms of the phase space elements and mean occupation numbers $n_{x,p_{\bot}}=N_{x,p_{\bot}}/G_{x,p_{\bot}}$ of each quantum state.  By choosing the cell sizes small we convert the sums into integrals over the continuum variables and then vary the entropy under the two constraints of light-cone momentum conservation and total transverse energy:
\begin{eqnarray}
g  L_{\bot}^2\int \frac{d^2 p_{\bot}}{(2 \pi)^2} \int \frac{dx}{x} 
x  n(x, p_{\bot})&=&1\\
g  L_{\bot}^2\int \frac{d^2 p_{\bot}}{(2 \pi)^2} \int \frac{dx}{x} 
p_{\bot} \ n(x, p_{\bot})&=&\langle E_{\bot}\rangle.
\end{eqnarray}
The variation gives the LCP distribution function $n(x,p_{\bot})$ as maximum-entropy distribution:
\begin{equation}
n(x,p_{\bot})=
\frac{1}{e^{\frac{p_{\bot}}{\lambda}+x w }-1}.
\end{equation}
For details we refer to Ref.~\cite{Pirner:2011ab}.

Light-cone momentum conservation serves as a constraint for the light-cone plasma distribution. Unlike in $e^+e^-$ collisions it cannot be expected that the center-of-mass energy is fully available for particle production. The fraction of the total center-of-mass energy available for particle production is denoted as $K$ in this paper. Proton-proton collisions at $\sqrt{s}=\unit[0.2]{TeV}$ are described by $K \approx 0.5$ \cite{Pirner:2011ab}. For pp collisions at $\sqrt{s}=\unit[7]{TeV}$ we use $K \approx 0.3$.

The LCP distribution together with the measure generate a gluon rapidity distribution of the following form:
\begin{equation}
\frac{d N}{dy d^2p_{\bot}}= \frac{g L_{\bot}^2}{(2 \pi)^2} 
\frac{1}{\exp\left[p_{\bot}\left(\frac{1}{\lambda}+\frac{w e^{|y|}}{K \sqrt{s}}\right)\right]-1}.
\end{equation}
This is the connection of the maximum-entropy distribution with the semi-inclusive cross section.  The phenomenological description of the multiplicity distribution has three parameters $L_{\bot}^2,\lambda$ and $w$.  The parameter $\lambda$ plays the role of an effective transverse ``temperature''. The ``softness" $w$ determines how small the mean $x$ becomes. With increasing center-of-mass energies, we expect that the effective transverse temperature $\lambda$ and $w$ increase: The collision becomes ``hotter'' and the particle distributions ``softer''.  The effective transverse temperature $\lambda$ is calculated from the mean transverse momentum which is equal to the ratio of the transverse energy and multiplicity in one hemisphere.

To compare with experiment we assume parton-hadron duality and  saturate the hadronic reaction products by pions such that the charged hadrons make up $2/3$ of the total multiplicity. In addition, we convert the multiplicity as a function of rapidity into a function of pseudorapidity using the Jacobian depending on the pion mass.
In reference \cite{Pirner:2011ab} we have determined the best values of $L_{\bot}$,$\lambda_{pp}$ and $w_{pp}$  for pp  collisions at cm energies of $\sqrt{s}=0.2\TeV$ and $\sqrt{s}=2.76\TeV$; see Table~1.
\begin{table}[h]
\begin{tabular}{c|c|c|c|c|c}
$\sqrt{s}$ & $L_\perp$ & $\lambda_{pp}$ & $w_{pp}$ & $\langle p_\perp^2
\rangle$ & $dN_\mathrm{ch}/d\eta$   \\
(TeV) & (fm) & (GeV) & & (GeV$^2$) &  \\
\hline
0.20 & 1.34 & 0.183 & 3.44  & 0.12 & 2.20   \\
2.76 & 1.28 & 0.252 & 6.81  & 0.23 & 4.56   \\
\end{tabular}
\caption{The first four columns give the cm energy $\sqrt{s}$, the transverse size $L_{\bot}$, the effective transverse temperature  $\lambda_{pp}$ and the softness $w_{pp}$ of the  pp light-cone distributions. The next columns contain the mean transverse momentum squared of gluons $\langle p_\perp^2\rangle$ (GeV$^2$) and the rapidity distribution $dN_\mathrm{ch}/d\eta$ at $\eta=0$ of charged particles.}
\end{table}
It is noteworthy that the statistical weights  from the spins and colors of gluons correctly reproduce the magnitude of the inclusive cross sections. 

In general, one would have to split the parton content of the proton into quarks and gluons such that the gluon momentum fractions do not integrate up to unity. The quarks and antiquarks have to be described by Fermi statistics.  The individual momentum fractions will then depend on the effective saturation scale of the collision. It is clear that not only the momentum distribution of the LCP, but also the relative amount of quarks and gluons in this description will be different from an equilibrated three-dimensional thermal quark-gluon plasma. With increasing energy the reaction products become more and more gluonic. This would also show up in the particle content after hadronization. Detailed consequences of such a more accurate description have to be explored theoretically and experimentally. Here in this paper we stick to the approximation where gluons are the only constituents of the proton and make up the full momentum sum rule.

\section{Transverse-Momentum Broadening in Nucleus-Nucleus Collisions}
To extend the calculation to nucleus-nucleus collisions we assume that the  multiplicity in the AA collision is proportional to the  number of participating nucleons $N_\mathrm{part}$:
\begin{equation}
\frac{d N_{ch}^{AA}}{d\eta d^2p_{\bot} } =N_\mathrm{part}\frac{2}{3} 
\sqrt{1- \frac{m_\pi^2}{m_\bot^2 \cosh^2(y)}} 
\frac{d N(\langle p_{\bot} \rangle)}{dy d^2p_{\bot}}.
\end{equation}
 
However, this is not sufficient. It is important to take into account the increase of the mean transverse momentum $\langle p_{\bot}\rangle$ with the number of participants. The initial
parton distributions in the projectile nucleus will be broadened by the interaction with the nucleons in the target nucleus and vice versa. This broadening will be calculated now.

Transverse-momentum broadening results from coherent rescatterings with vanishing small momentum transfers.  Averaging the differential cross section with the acquired transverse momentum yields the transport coefficient which is defined as the mean transverse momentum squared times the cross section, i.e., $\langle\sigma p_{\bot}^2\rangle$.  This form has been tested in electron-nucleus and proton-nucleus collisions \cite{Domdey:2008aq}. The resulting mean $\Delta p_{\bot}^2$ is also known as saturation scale $Q_s^2$ \cite{McLerran:1993ni,*McLerran:1993ka}.  It can be derived from the dipole nucleon cross section as follows.

In the eikonal approximation, the ejected high-momentum parton moves on a classical trajectory with impact parameter $\vec{b}$ and picks up a non-Abelian phase factor
$V(\vec{b})$ in the background gauge field generated by the nucleon:
\begin{eqnarray}
V(\vec{b})&=&{\cal P}{\,}\exp\left[\mathrm{i}\,g\int_{-\infty}^{+\infty}dx^\mu\, A_\mu(x)\right]\,.
\end{eqnarray}
Here $V(\vec{b})$ is the Wilson line of the parton with impact parameter $\vec{ b}$ relative to the proton. We use the notation $A_\mu\equiv A_\mu^a\,t^a$, where
$t^a$'s are the generators of the group SU($N_c$) in the fundamental representation. The differential cross section to produce a parton with transverse momentum $\vec{p}_\perp$ is
given by projecting the eikonal phase onto $\vec{p}_\perp$ and by taking the modulus of the amplitude integrated over all possible impact parameters
\begin{eqnarray}
\lefteqn{\frac{d\sigma}{d^2p_\perp}=} & & \\
& & \quad \frac{1}{(2\,\pi)^2}\int d^2b\,d^2b^\prime 
 {\rm e}^{i\,\vec{ p}_\perp(\vec{b}-\vec{b}^\prime)}{\,}\frac{1}{N_c}\left\langle{\rm Tr}\left[V^\dagger(\vec{b}^\prime)\,V(\vec{b}) \right]\right\rangle\,. \nonumber
 \label{eqn:PtCrossSec}
\end{eqnarray}
Hence, a fake dipole of size
$\vec{r}_{\bot}=\vec{b}-\vec{b}^\prime$ is constructed from the
ejected parton in the $V$ amplitude and in the
$V^\dagger$-amplitude. Their trajectories  are displaced from each
other by the distance $r_{\bot}$. The expectation values of the
Wilson lines have to be evaluated with respect to the target
ground state.
In the dipole model, the total cross section for the interaction
of a dipole of size $\vec{r}_{\bot}$ with a target nucleon is given by
\begin{eqnarray}
\lefteqn{\sigma_{dN}(\vec{r}_{\bot}) =} & & \nonumber \\
& & \quad 2\, \int d^2b \left(1-\frac{1}{N_c}\left<{\rm Tr}\left[V^\dagger(\vec{b}+\vec{r}_\perp)\,V(\vec{b})\right]\right>\right)\,.
\end{eqnarray}
We define the quantity $\left<\sigma p_\perp^2\right>$ as the integral over transverse momentum $d^2p_\perp$ of the differential cross section multiplied by $p_\perp^2$.
One sees that
$\left<\sigma p_\perp^2\right>$ is related to the dipole nucleon cross section:
\begin{eqnarray}
\left<\sigma p_\perp^2\right>
&\equiv& \int d^2p_\perp \frac{d\sigma}{d^2p_\perp}p_{\perp}^2\nonumber\\
&=&\frac{1}{(2\,\pi)^2}\int d^2p_\perp \int d^2b\,d^2r_\perp \left(-\nabla_{\bot}^2
{\rm e}^{i\vec{ p}_\perp\vec{ r}_{\bot}}\right){\,} \nonumber\\
&&\times \frac{1}{N_c}\left\langle{\rm Tr}\left[V^\dagger(\vec{b}+\vec{r}_\perp)\,V(\vec{b}) \right]\right\rangle\nonumber\\
&=&\frac{1}{2}\left. \nabla_{\bot}^2 \sigma_{dN}(\vec{r}_{\bot})\right|_{r_\perp=0}.
\label{broad}
\end{eqnarray}
This expression confirms the result derived in \cite{Johnson:2000dm}.
To obtain the transverse momentum broadening of gluons one has to consider the dipole-nucleon cross section as a function of dipole size $r_{\bot}$ and dipole-nucleon cm energy $\sqrt{\hat s}$.

Let us first consider $(3 \bar 3)$ dipoles. In the parametrization of Ref.~\cite{Kopeliovich:1999am} which was adjusted to data on deep-inelastic scattering (DIS) and photoproduction, the fast rise of the dipole cross section for small dipoles at high energies is included in the energy dependence of $r_0(\hat s)$:
\begin{equation}
\sigma_{dN}(\vec{r}_{\bot})=\sigma_0(\hat s)\left[1-\exp\left(-\frac{\vec{r}_\perp{}^2}{r_0^2(\hat s)} \right) \right]\,
\label{kst}
\end{equation}
with  
\begin{equation}
\sigma_0(\hat s)=23.6\,\left(\frac{\hat s}{s_0}\right)^{0.08}\left(1+\frac38
\frac{r_0^2(\hat s)}{0.44\,\mathrm{\fm}^2}\right)\,\mathrm{mb}
\end{equation}
and 
\begin{align}
r_0(\hat s)&=0.88\,(\hat s/s_0)^{-0.14}\,\mathrm{\fm},\\
s_0&=1000\,\mathrm{GeV}^2.
\end{align}
For gluons the transport-coefficient is modified by the color factor $9/4$: 
\begin{eqnarray}
\langle \sigma p_{\bot}^2(\hat s) \rangle_g&=&\left({9\over4}\right)\frac{1}{2}\left. \nabla_{\bot}^2 \sigma_{dN}(\vec{r}_{\bot})\right|_{r_\perp=0}\\
                             &=&\left({9\over4}\right)\frac{2 \sigma_0(\hat s)}{r_0^2(\hat s)}.
\end{eqnarray}

In pp collisions gluons are liberated with their intrinsic momenta described by the light-cone distribution $n(x,p_{\bot})$. In nuclear collisions this distribution is modified.  We will calculate the effect of multiple scattering on the transverse temperature $\lambda_{AA}$ and then insert this new transverse temperature in the universal light-cone distribution to determine the central rapidity distribution.  The effective cm energy $\hat s$ for multiple scattering has to be calculated from the collision of a transverse gluon with $(E_{\bot},p_{\bot},0)$ colliding with a proton in the opposite nucleus with four-momentum $(\sqrt{s_{NN}}/2,0,\sqrt{s_{NN}}/2)$:
\begin{eqnarray}
\hat s  = p_{\bot} \sqrt{s_{NN}}.
\end{eqnarray}

In the following we denote the averaging over $\hat s$ with $n(x,p_{\bot})$ by a bar.  For $n(x,p_{\bot})$ we use the parameters $\lambda_{pp}$ and $w_{pp}$ of the corresponding light-cone gluon distributions from Table~1 in Sec.~II. These are the parameters which describe the multiplicity distributions for Relativistic Heavy Ion Collider (RHIC) and Large Hadron Collider (LHC) energies as shown in Ref.~\cite{Pirner:2011ab}. We obtain
\begin{eqnarray}
\overline{\langle \sigma p_{\bot}^2 \rangle_g}&=&\frac {\sum \langle \sigma p_{\bot}^2( \hat s)\rangle_g n(x,p_{\bot})}{\sum n(x,p_{\bot})}\\
&=& 
\left\{
  \begin{array}{l l}
    11.26 & \text{for } \sqrt{s_{NN}}=\unit[0.2]{TeV}\\
    21.15 & \text{for } \sqrt{s_{NN}}=\unit[2.76]{TeV}.\\
  \end{array} 
\right.
\label{transport}
\end{eqnarray}

\begin{figure}[ht]
\centering
\includegraphics[width=0.95\linewidth]{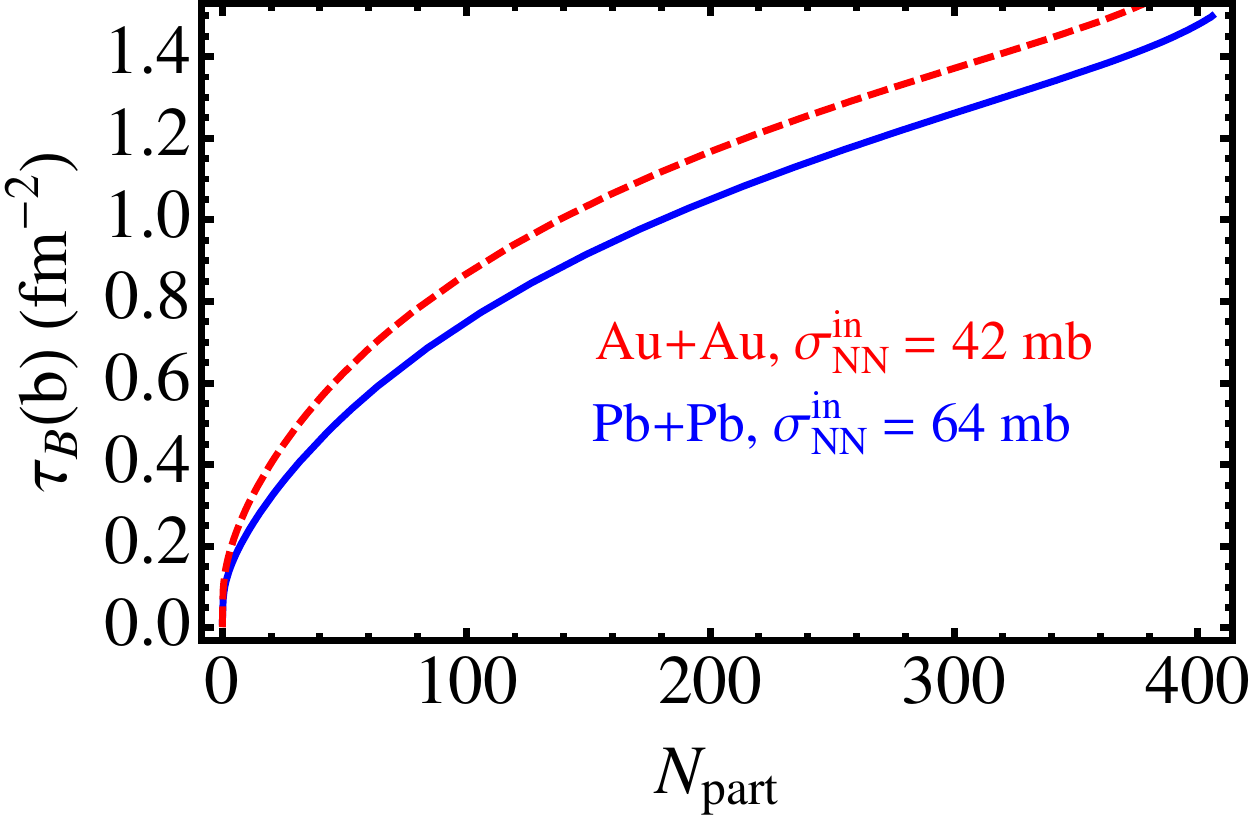}
\caption{(Color online) The average profile functions $\tau(b)$ [see Eq.~(\ref{eq:tau})] for Au+Au at $\sqrt{s_{NN}}=0.2 \TeV$ (dashed curve) and Pb+Pb collisions at $\sqrt{s_{NN}}=2.76\TeV$ are shown as functions of the number of participants. The number of participants increases for more central collisions. The functional dependence of the resulting mean profile functions varies because the relation between impact parameter and number of participants depends on cm energy.}
\label{fig:tb}
\end{figure}

In the second step we multiply the mean transverse momentum in a collision with the average profile function. This gives the mean $\Delta p_{\bot}^2$ of gluons:
\begin{eqnarray}
 \Delta p_{\bot}^2=\overline{\langle \sigma p_{\bot}^2\rangle_g}\tau_B(b).
\label{40}
\end{eqnarray}

We calculate the averaged profile function $\tau_B(b)$ now:
The profile functions $T_A(b)$ and $T_B(b)$ of the colliding nuclei A and B are obtained from the nuclear densities with Woods-Saxon distributions: 
\begin{align}
T_A(b) &= \int \rho_A(\sqrt{b^2+z^2})\,dz, \\
 \rho_A(r) &=\frac{\rho_0}{1+\exp[(r-R)/a]}.
\end{align}
We use $R = \unit[6.62 \,(6.38)]{fm}$ and $a = \unit[0.54]{fm}$ to describe the Pb (Au) nucleus, respectively \cite{De_Jager:1987qc}. 


The number of participants (for a definition see e.g. Ref.~\cite{Miller:2007ri}) depends on the inelastic nucleon-nucleon cross section, for which we take the values $\sigma_{NN}^\mathrm{in}(\unit[0.2] {TeV})=42\mb$ and $\sigma_{NN}^\mathrm{in}(\unit[2.76]{TeV})=\unit[64]{mb}$ from Ref.~\cite{Adler:2004zn,*Aamodt:2010cz}. In AB collisions, the total number of participants, $N_\mathrm{part}(b)$, equals the number of participants in nucleus A and in nucleus B which explains the factor two in the equation below for symmetric AA collisions:
\begin{equation}
N_\mathrm{part}(b) = 2 \int d^2s\, T_A(\vec s) \left\{1 - \exp\left[- T_B(\vec b -\vec s)\sigma_{NN}^\mathrm{in}\right]\right\}.
\end{equation}
The averaged profile function $\tau_B(b)$ of nucleus B is then obtained by using the number of participant nucleons in A as weight function: 
\begin{equation}
\tau_B(b) = \frac{\int d^2s\, T_A(\vec s) \{1 - \exp[- T_B(\vec b -\vec s) \sigma_{NN}^\mathrm{in}]\}T_B(\vec b-\vec s)}{\int d^2s\, T_A(\vec s)\{1- \exp[- T_B(\vec b -\vec s) \sigma_{NN}^\mathrm{in}]\}}.
\label{eq:tau}
\end{equation}

Both the number of participants and the average profile function are functions of the nucleus-nucleus impact parameter $b$. In experiment, the number of participants is used as a measure of the centrality of the collision. Therefore we eliminate the impact parameter $b$ and represent the mean profile as a function of the number of participants, i.e.\ centrality. The dependence of the mean profile function as a function of the number of participants is rather well described by a power law:
\begin{equation}
\tau_B(N_\mathrm{part}) \approx n_0 N_\mathrm{part}^{a}.
\end{equation}

The corresponding values for the parameters are $n_0=0.1155 \fm^{-2}$, $a=0.435$ for $A=197$, and $n_0=0.0775 \fm^{-2}, a=0.491$ for $A=208$. A plot of the numerical profiles for both nuclear collisions is shown in Fig.~\ref{fig:tb}.

For an estimate of the mean path length traversed by a parton in the AA collision one has to divide this profile density by the mean nuclear density $\rho_0 \approx 0.17 \fm^{-3}$. One obtains a mean path length of approximately 10 fm for central collisions.  Owing to the larger inelastic pp cross section at the LHC the $b$ distribution of the number of participants extends to larger impact parameters for the LHC. Therefore, the weight of shorter trajectories is increased at the LHC. This explains the difference of the mean profile functions in Fig.~\ref{fig:tb}.

\begin{figure}[ht]
\centering
\includegraphics[width=0.95\linewidth]{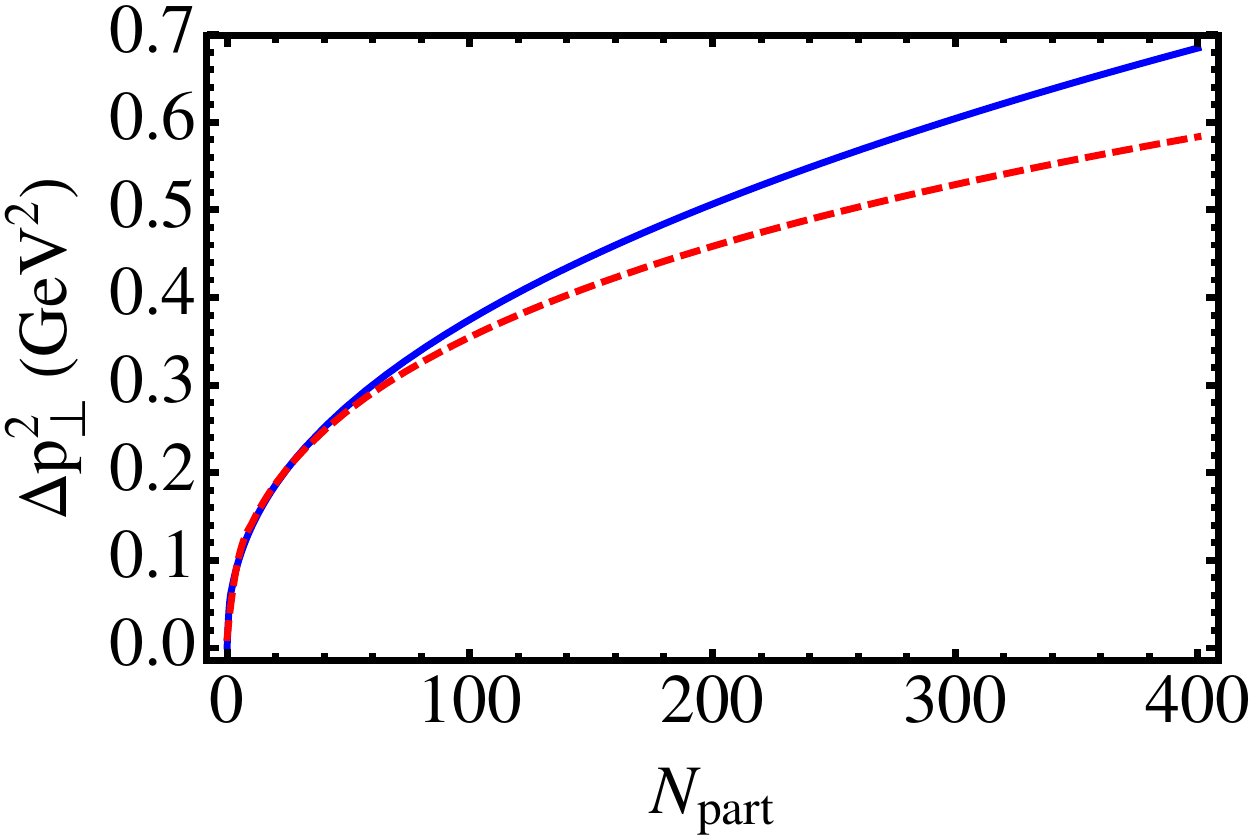}
\caption{(Color online) The average broadening $\Delta p_{\bot}^2$ of a gluon for Au+Au at $\sqrt{s_{NN}}=0.2\TeV$ is shown as functions of the number of participants. The number of participants increases for more central collisions. The solid blue line represents the result from the multiple scattering calculation. The dashed red line is the result from mutual boosting and shadowing.}
\label{fig:dpt2}
\end{figure}

Having described the relevant steps we can now calculate the transverse momentum broadening for gluons in nucleus-nucleus collisions for the two different energies $\sqrt{s_{NN}}=0.2$ TeV and $\sqrt{s_{NN}}= \unit[2.76]{TeV}$. Neglecting the small differences of the average profile functions between Au-Au and Pb-Pb at the different energies we see that the transport coefficient determines the energy dependence of momentum broadening. This coefficient increases by a factor of two between RHIC and LHC, therefore the mean transverse momentum broadening increases by the same factor. In Figs.~\ref{fig:dpt2} and ref{fig:dpt1} the thick (blue) lines represent the mean value of $\Delta p_{\bot}^2$ of gluons as a function of the number of participants in the AA collision for RHIC and LHC energies calculated using the dipole scattering cross section. The two curves have almost identical shapes because the profile functions in Au-Au and Pb-Pb collisions are nearly equal; see Fig.~1.  The shape of the curves is determined by the geometry of the nuclear collisions.
 
\begin{figure}[ht]
\centering
\includegraphics[width=0.95\linewidth]{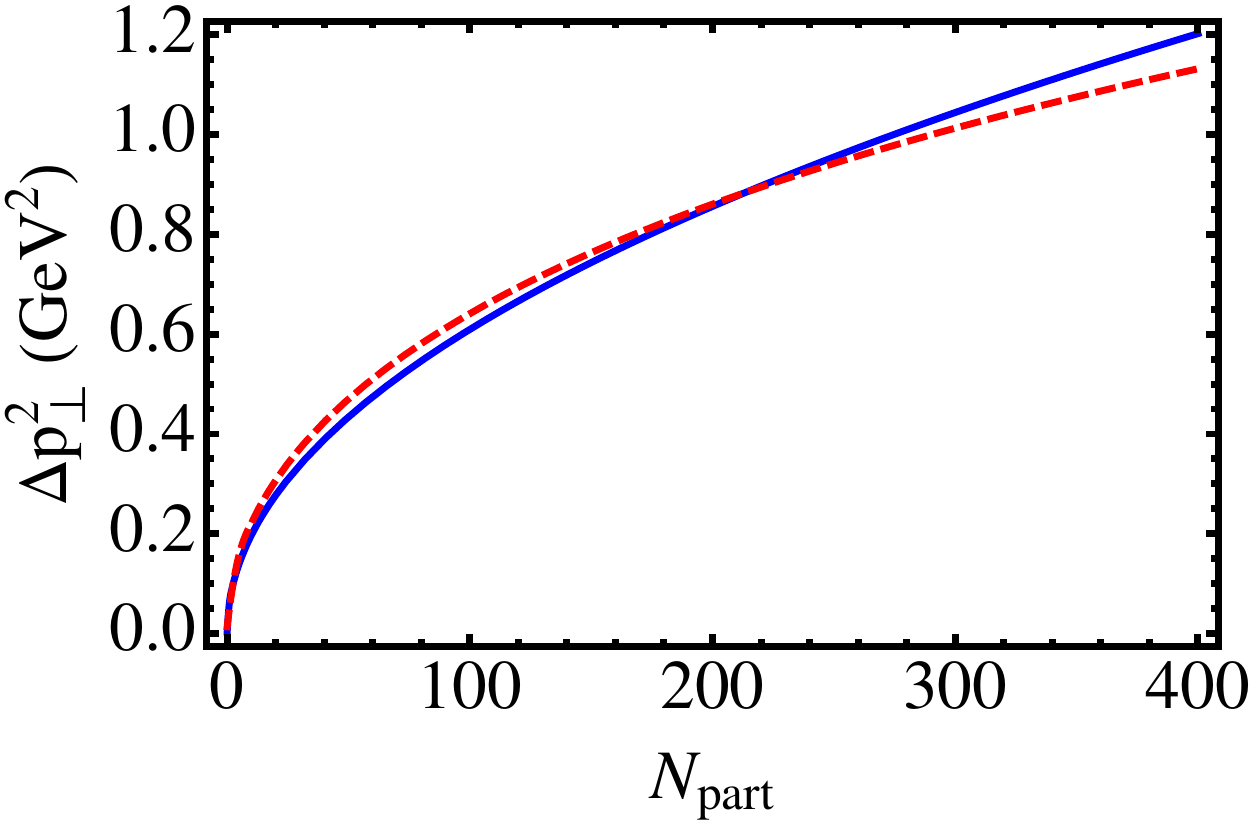}
\caption{(Color online) The average broadening $\Delta p_{\bot}^2$ of a gluon for Pb+Pb at $\sqrt{s_{NN}}=2.76\TeV$ is shown as functions of the number of participants. The number of participants increases for more central collisions. The solid line represents the result from the multiple scattering calculation. The dashed red line is the result from mutual boosting and shadowing.}
\label{fig:dpt1}
\end{figure}

\section{Mutual Boosting of Saturation Scales and Shadowing}
In proton nucleus scattering the transverse momentum of a gluon in the proton propagating through the target nucleus B increases by the value $\Delta p_{\bot}^2$, which depends on the profile $\tau_B(b)$ of the target. The resulting momentum broadening also defines the saturation scale $Q_{sB}^2$ of the intrinsic gluon distribution of the target nucleus B \cite{Kopeliovich:2010aa}:
\begin{equation}
Q_{sB}^2= \Delta p_{\bot}^2.
\end{equation}

This picture of broadening strictly applies only when we consider nucleus-nucleus collisions as a superposition of proton-nucleus collisions.  In nuclear collisions, however, multiple interactions enhance the higher Fock components in the wave functions of the participating nucleons, thereby increasing the saturation scale compared with NA collisions \cite{Kopeliovich:2010mx}. The saturation scales both in the target and projectile nucleus are boosted to higher values. Intuitively, a nucleon Fock state distribution in the nucleus B is affected by multiple interactions, which enhances the contribution of multi-parton components. Thus, the gluon density in the bound nucleons in B increases at small $x$, which in turn leads to a more intensive interactions of these nucleons with partons from the nucleus A, i.e.\ enhances broadening of partons propagating through B.  This effect of mutual boosting of saturation scales has been described first in \cite{Kopeliovich:2010mx} and was subsequently discussed in Ref.~\cite{Kopeliovich:2010nw,Kopeliovich:2011zz}.  On the other hand, multiple interactions at high energies are subject to shadowing, which works in the opposite direction, suppressing the magnitude of broadening and saturation scales \cite{Kopeliovich:2010aa}.

For high energies the dipole-nucleon interaction is dominantly given by two-gluon exchange. The transport parameter derived in the previous section can then be related to the gluon structure function at a certain Bjorken $\bar x$ and scale $Q_0^2$.
In order to make this perturbative picture coincide with the nonperturbative model based on the dipole cross section of Sec.~III , we determine the infrared scale in such a way that the two-gluon exchange picture on the proton reproduces the nonperturbative result
\cite{Kopeliovich:2010mx}, 

\begin{equation}
\frac{2}{9}\overline{\langle\sigma p_{\bot}^2\rangle_g}=\frac{\pi^2}{3}\alpha_s(Q_0^2) \overline{\bar x g_N(\bar x,Q_0^2)},
\end{equation}

with

\begin{equation}
\bar x=\frac{Q_0^2}{\hat s}=\frac{Q_0^2}{p_{\bot} \sqrt{s_{NN}}}.
\end{equation}

Notice that both sides of the above equation depend on the transverse momentum of the gluon after its liberation on mass shell.  We average the left and right hand sides of this equation over all possible $p_{\bot}$ values using the light-cone distribution $n_{pp}(x,p_{\bot})$ generated in pp collisions with the parameters of Table.~I to be consistent with Eq.~(\ref{transport}).  For scale setting we need a next-to-leading order (NLO) distribution. Most distribution functions, such as the MSTW distribution \cite{Martin:2009iq} in NLO are defined for $Q^2>1\GeV^2$ and do not allow low enough virtualities to set the infrared scale $Q_0^2$.  The NLO-GRV parametrization \cite{Gluck:1998xa}, however, extends to low virtualities $Q^2 \approx 0.4\GeV^2$ and low $x$. We can use this distribution to determine the infrared scale $Q_0^2$. We find the following solutions of the selfconsistency equations for RHIC and LHC energies:
\begin{eqnarray}
Q_0^2(0.2\TeV)=0.35\GeV^2\\
Q_0^2(2.76\TeV)=0.41\GeV^2.
\end{eqnarray}

The effective temperature and the softness parameters given in Table~I influence the infrared scale. It is therefore not unreasonable that the infrared scales obtained in this way are close to the $p_{\bot}^2$-scales of the multiplicity distributions at the same energies; see Table~I. Our values for the infrared scales may in principle differ from the pp saturation scale associated with the color glass parametrizations; see Ref.~\cite{Dumitru:2011wq}. We obtain a slightly smaller infrared scale $Q_0^2=0.35\GeV^2$ than the saturation scale $Q_s^2=0.462\GeV^2$ at RHIC energy. Obviously, the sizes of the infrared scales only marginally justify the use of perturbative QCD. It may be useful to consider the saturation of the nonintegrated gluon structure function in a nonperturbative formulation (see Ref.~\cite {Shoshi:2002fq}) based on string-string interactions. We will reserve such a work to further studies.

The boosting effect alone  overestimates the magnitude of broadening. It must be considered together with the effect of nuclear shadowing which reduces the nuclear gluon distribution compared with the pure sum of nucleonic gluon distributions at small $x$. The nuclear gluon distribution does not linearly increase with the nuclear profile function. A simple parametrization of the results in Fig.~2  of reference \cite{Kopeliovich:2010aa} has the following form:
\begin{eqnarray}
\label{eq:gb}
g_B(x)&=&g_N(x) \tau_B(b) \left\{1- \alpha \ln[ \tau_B(b)/\tau_0 ]\right\}\\
      &=&g_N(x) \tilde \tau_B(b).
\end{eqnarray}

The parameter $\alpha$ is only weakly dependent on energy. In the above reference the authors find for RHIC $\alpha \approx 0.15$ and for LHC $\alpha \approx 0.17$ with $\tau_0=0.1 \fm^{-2}$.  This reduction enters the differential equations for the the saturation scale and $p_{\bot}$ broadening. The increase $dQ_{sB}^2$ of the saturation scale in nucleus B given by the $p_{\bot}$ broadening of a parton of nucleus A going through nucleus B rises with the increase of the modified profile function $d \tilde \tau_B(b)$.  A similar equation holds for $dQ_{sA}^2$: 
\begin{eqnarray} 
\frac{dQ_{sB}^2}{d \tilde \tau_B(b)} &=&\frac{3 \pi^2}{2}\alpha_s(Q_{sA}^2+Q_0^2) \overline{\bar x_A g_N (\bar x_A,Q_{sA}^2+Q_0^2)} \quad
\label{48}\\
\frac{dQ_{sA}^2}{d \tilde \tau_A(b)} &=&\frac{3 \pi^2}{2}\alpha_s(Q_{sB}^2+Q_0^2) \overline{\bar x_B g_N(\bar x_B,Q_{sB}^2+Q_0^2)}. \quad \label{49} 
\end{eqnarray}

Since the saturation scale $Q_{sB}^2$ of nucleus B increases with the saturation scale $Q_{sA}^2$ of nucleus A, the authors of Ref.~\cite {Kopeliovich:2010mx} have called this phenomenon mutual boosting. With increasing saturation scale also the associated Bjorken variable $\bar x_A$ in the gluon structure functions changes ($\bar x_B$ is defined in the same way with index $A \rightarrow B$):
\begin{equation}
\bar x_{A}=\frac{Q_0^2+Q_{sA}^2}{\hat s}.
\end{equation}
The right hand sides of the differential equations depend on the $\bar x$ values of the gluon structure functions. Since the structure functions become larger at smaller $\bar x$, the increase of broadening is linked with increasing $p_{\bot}$ of the incoming gluon in the same way as the dipole cross where increasing $p_{\bot}$ means higher dipole energy and larger transport parameter.

These differential equations have to be calculated with the initial conditions that $Q_{sA}^2=Q_{sB}^2=0$ for $\tau_A(b)=\tau_B(b)=\tau_0$ [see Eq.~(\ref{eq:gb})], which guarantee that the results for peripheral nuclear collisions coincide with the results of Sec.~III obtained without boosting. We prefer these differential equations to the equivalent bootstrap equations of Ref.~\cite{Kopeliovich:2010mx}, because they allow us to include the averaging in an easier way. In the case of symmetric AA collisions both equations coincide, because $\tau_B(b)= \tau_A(b)$ and $Q_{sA}^2=Q_{sB}^2$.  In Fig.~\ref{fig:dpt2} and Fig.~\ref{fig:dpt1} the dashed lines show the dependence of the perturbative saturation scale $Q_{sA}^2$ on centrality expressed by the number of participants.  Comparing these values with the corresponding values for $\Delta p_{\bot}^2$ from the calculation in Sec.~III, one sees that for RHIC and LHC the mutual boosting of saturation scales gives approximately the same results.  Shadowing is very important to control the amount of mutual boosting, which otherwise would overshoot the calculation from multiple scattering by a factor of two.

\section{$dN/d\eta$ distributions in AA collisions} 
Having a theoretical calculation of gluon $p_{\bot}$ broadening or equivalently of the saturation scales in nucleus-nucleus collisions, we can calculate the change of the transverse temperature $\lambda$ and the softness as a function of centrality.  Because of the well defined shape of the LCP distributions we can relate the calculated mean $p_{\bot}$ values in AA collisions to $dN/d\eta$ distributions.
  
To describe the momentum broadening of pions one has to convert the information about gluons obtained in Secs.~III and IV into pion spectra. Since the transverse momentum broadening of gluons described in the nuclear multiple scattering picture is intrinsically higher than for pions, one cannot use here parton-hadron duality. To take into account the fragmentation of the gluons into pions, we use the fragmentation function of Ref.~\cite{Albino:2005me} at the starting scale of $Q_0^2=2\GeV^2$. The mean $\langle z^2\rangle_{\pi/g}$ relates the pion momentum broadening to the gluon broadening: 
\begin{eqnarray}
D_{\pi/g}(z)&=&429 z^2(1-z)^{5.82}\\
\langle z^2\rangle_{\pi/g}&=&0.11\;.
\end{eqnarray}
We use the result of Eq.~(\ref{40}) of Sec.~III:
\begin{equation}
\Delta p_{\bot,\pi}^2=\overline{\langle\sigma p_{\bot}^2\rangle_g}\tau_B(b) \langle z^2\rangle_{\pi/g}.
\end{equation}
Mutual boosting explained in Sec.~IV gives saturation scales which have been identified with gluon momentum broadening. The saturation scales $Q_{sA}^2$ obtained from the solution of the differential equation  Eqs.~(\ref{48})-(\ref{49}) can also be converted into pion momentum broadening:
\begin{equation}
\Delta p_{\bot,\pi}^2=Q_{sA}^2\langle z^2\rangle_{\pi/g}.\\
\end{equation}
The total mean pion transverse momentum is given by the sum of the intrinsic pp mean transverse momentum $\langle p_{\bot}^2\rangle$ given in Table~I and the acquired $\langle\Delta p_{\bot}^2\rangle_{\pi}$ above:
\begin{equation}
\langle p_{\bot,tot}^2\rangle=\langle p_{\bot}^2\rangle+ \Delta p_{\bot,\pi}^2.
\label{eq:pttot}
\end{equation}

The universal form of the LCP distributions has been shown in Ref.~\cite{Pirner:2011ab} to describe well the rapidity and $p_{\bot}$ distributions in AA collisions. Since the shape of the distribution is fixed by the maximum-entropy principle one only has to modify the effective transverse temperature and softness parameters of the pp light-cone distributions to obtain the nuclear light-cone distributions. The effective transverse temperatures $\lambda_{AA}$ are related to the mean transverse momenta of gluons averaged over all rapidities  as in Table~I:

\begin{equation}
\lambda_{AA} \approx 0.53 \sqrt{\langle p_{\bot,\mathrm{tot}}^2\rangle} 
\label{eq:lambda_aa}
\end{equation}

We use the $x$ sum rule for fixed $L_{\bot}= \unit[1.28]{fm}$ taken from pp data (see Table~I) and the values of $\lambda_{AA}$ modified by momentum broadening, see Eqs.~\ref{eq:pttot} and \ref{eq:lambda_aa}. Then we can calculate the resulting softness parameters $w_{AA}$ for centralities from 0\% to 30\%. One finds the following residual parameters $w_{AA}$ for Pb-Pb collisions at $\sqrt{s_{NN}}= \unit[2760]{GeV}$ see Table~II.

\begin{table}[h]
\begin{tabular}{c|c|c|c|c}
Centrality     &$N_\mathrm{part}$ & $L_{\bot}$ & $\lambda_{AA}$ & $w_{AA}$  \\
             &                  & (fm)     & (GeV)          &                                    \\                 
\hline
20\%-30\% &  186   & 1.28 & 0.299  & 9.99  \\
10\%-20\% &  260   & 1.28 & 0.306  & 10.55 \\
0\%-5\% &  383   & 1.28 & 0.316  & 11.30 
\end{tabular}
\caption{For different centralities related to the number of participants in Pb+Pb collisions at $\sqrt{s_{NN}}=\unit[2760]{GeV}$   the table gives the LCP parameters. Given fixed sizes $L_{\bot}$, increasing effective temperatures $\lambda_{AA}$ from momentum broadening make the softness
parameters $w_{AA}$ increase. ($K=0.35$).}
\label{tab:Pb_Pb}
\end{table}

The form of the inclusive cross sections is given by the universal LCP distributions with these parameters.
\begin{equation}
\frac{d N_{ch}^{AA}}{d\eta d^2p_{\bot} } =\frac{N_\mathrm{part}}{2}\frac{2}{3} 
\sqrt{1 - \frac{m_\pi^2}{m_\bot^2 \cosh^2 y}} 
\frac{d N(\lambda_{AA},w_{AA})}{dy d^2p_{\bot}}
\label{eq:npart_scaling}
\end{equation}
with
\begin{equation}
\frac{d N}{dy d^2p_{\bot}}= \frac{g L_{\bot}^2}{(2 \pi)^2} 
\frac{1}{\exp\left(m_\bot(\frac{1}{\lambda_{AA}}+\frac{w_{AA} e^{|y|}}{K \sqrt{s_{NN}}})\right)-1}.
\end{equation}

\begin{figure}[!]
\centering
\includegraphics[width=\linewidth]{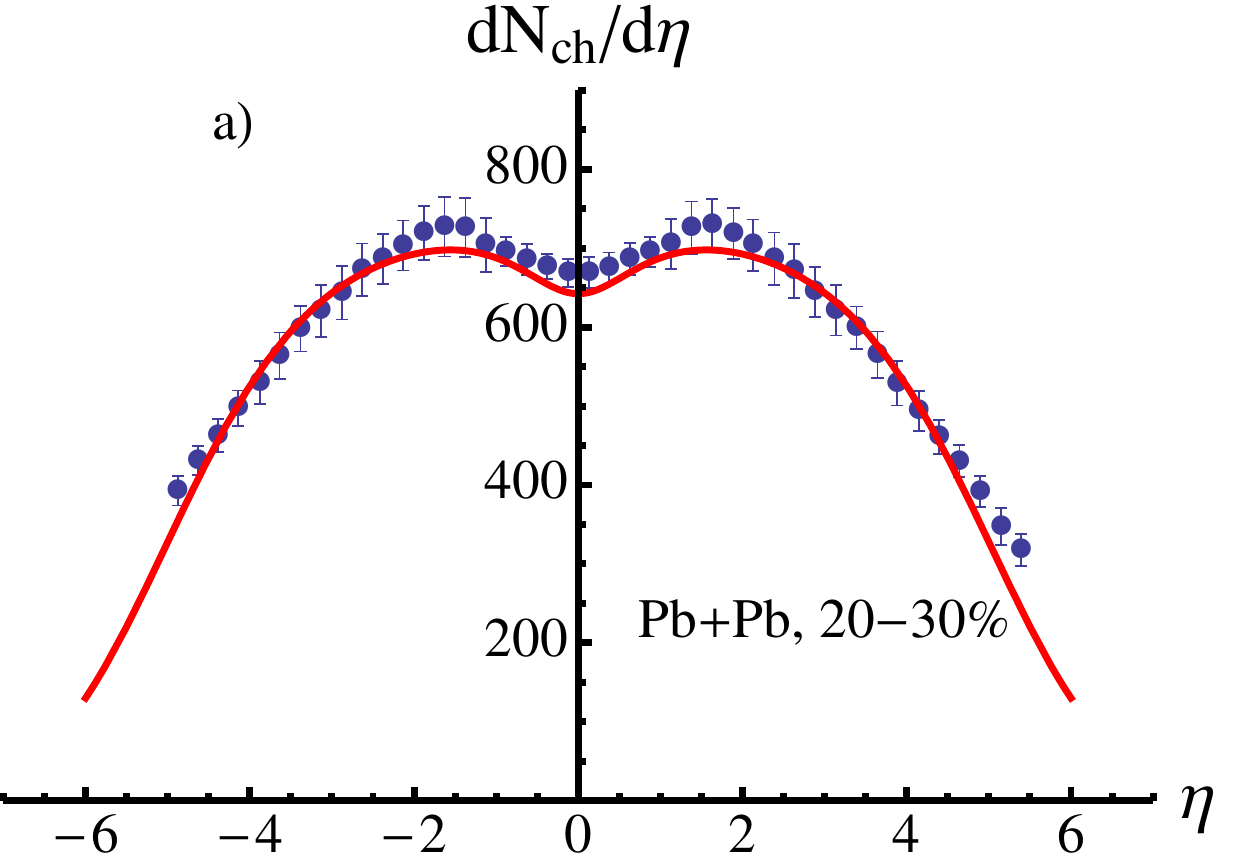}
\includegraphics[width=\linewidth]{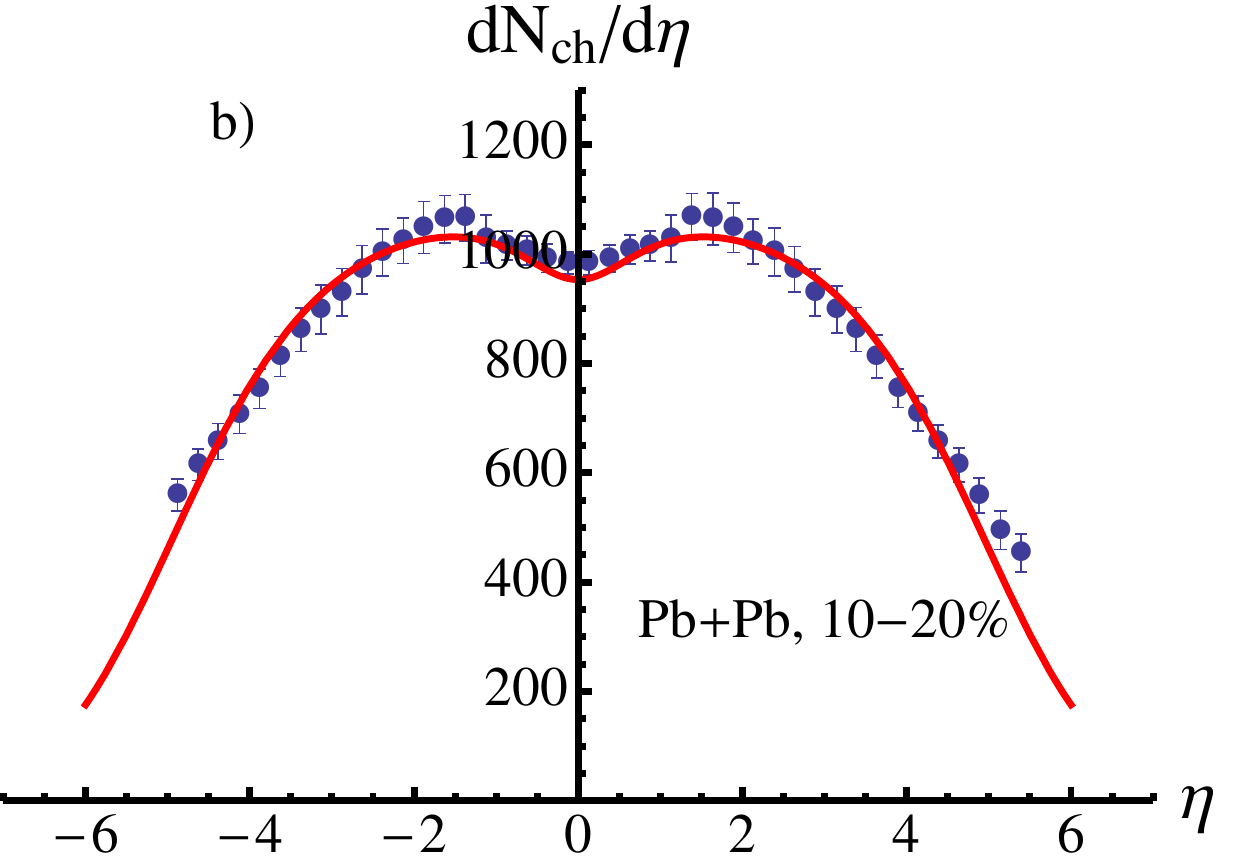}
\includegraphics[width=\linewidth]{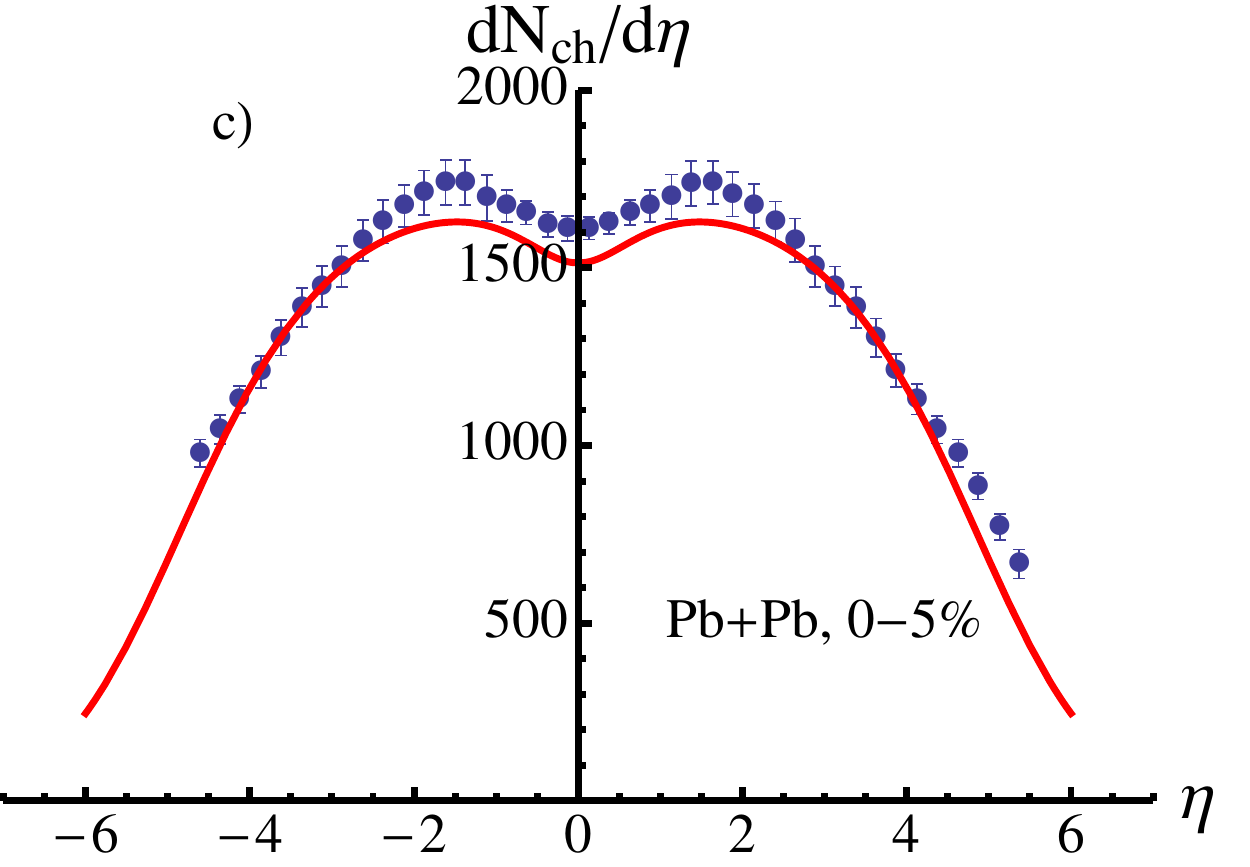}
\caption {(Color online) Preliminary charged particle pseudorapidity distributions from ALICE \cite{Gulbrandsen:2013iu} in Pb+Pb collisions at $\sqrt{s_{NN}} = \unit[2.76]{TeV}$ for the following centrality classes: (a) $20-30\,\%$, (b) $10-20\,\%$, and (c) $0-5\,\%$. Data are compared with the theoretical curves obtained from the light-cone plasma distributions ($K=0.35$)}
\label{fig:dnchdeta_heavy_ions}
\end{figure}

In Fig.~\ref{fig:dnchdeta_heavy_ions} we show a comparison of the full pseudorapidity distributions for the Pb-Pb collisions with preliminary ALICE data \cite {Gulbrandsen:2013iu}. The theoretical curves are predictions based on the calculated effective temperatures $\lambda_{AA}$ obtained from transverse momentum broadening. If one would fit the measured data, one would obtain a slightly larger $L_{\bot}$ for central collisions $L_{\bot} \approx \unit[1.31]{fm}$ which corrects the normalization. The width of the rapidity distribution is influenced by the $K$ factor which decreases with increasing energies.  The $K$ factor $K=0.7$ for 200 GeV collisions decreases to $K=0.35$ for 2760 GeV collisions, a decrease similar to that already seen in pp collisions.  For 200~GeV Au-Au collisions we obtain results very similar to the already published pseudorapidity distributions in \cite{Pirner:2011ab}.

\begin{figure}[ht]
\centering
\includegraphics[width=0.95\linewidth]{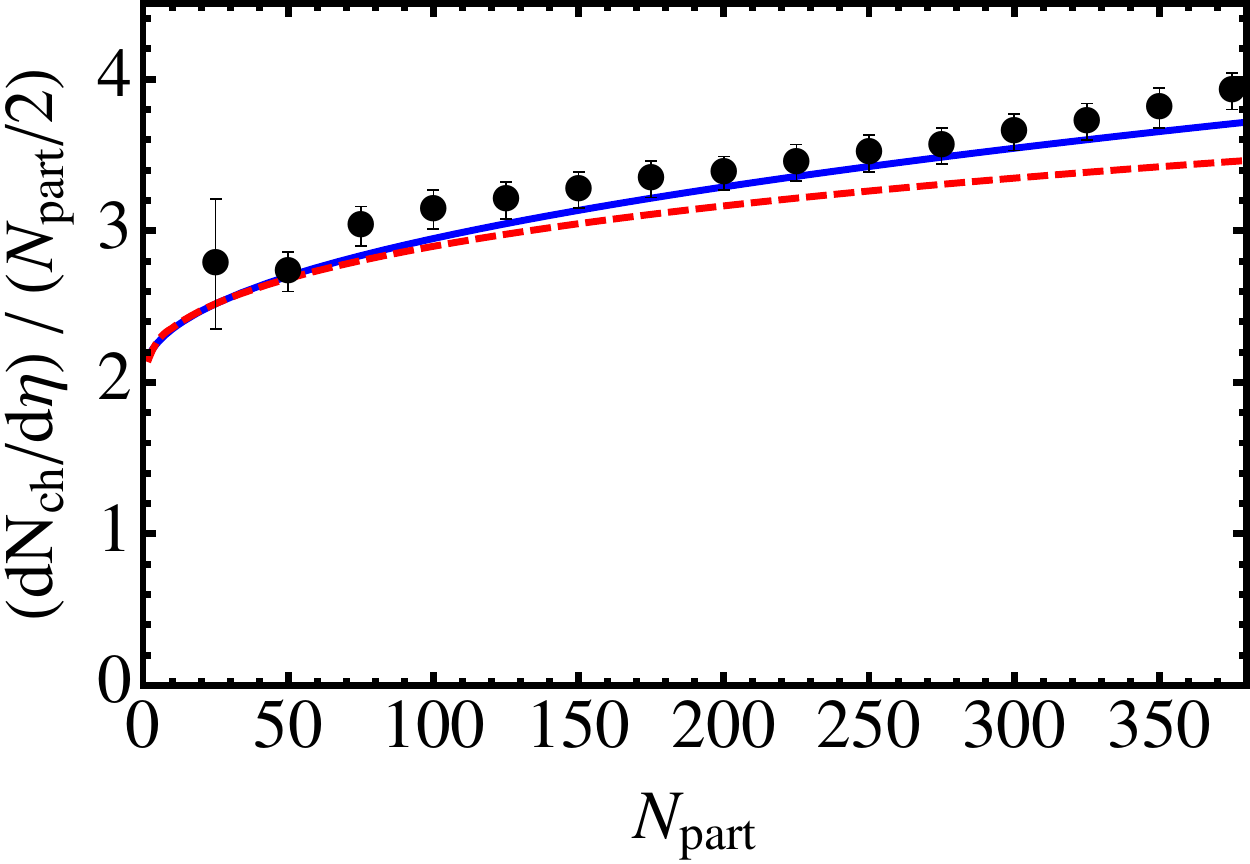}
\caption{(Color online) The multiplicity $dN_\mathrm{ch}/d\eta$ at $\eta=0$ for Au+Au at $\sqrt{s_{NN}}=0.2\TeV$ is shown as a function of the number of participants. The number of participants increases for more central collisions. The solid line represents the result from the multiple scattering calculation. The dashed line is the result from mutual boosting and shadowing. Data points represent an average of results from the BRAHMS, PHENIX, PHOBOS, and STAR experiments \cite{Adler:2004zn}.}
\label{fig:dNchdeta2}
\end{figure}

\begin{figure}[ht] 
\centering
\includegraphics[width=0.95\linewidth]{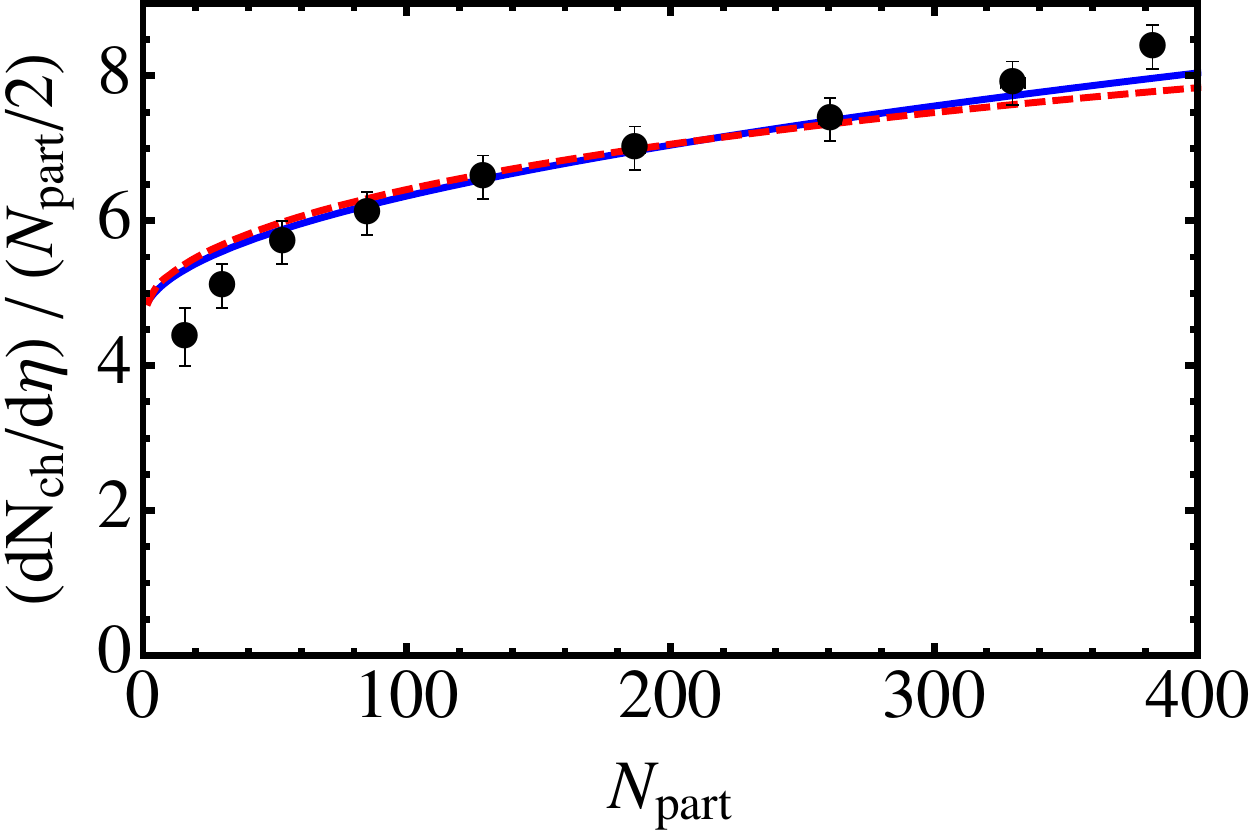}
\caption{(Color online) The multiplicity $dN_\mathrm{ch}/d\eta$ at $\eta=0$ for Pb+Pb at $\sqrt{s_{NN}}=2.76\TeV$ is shown as a function of the number of participants. The number of participants increases for more central collisions. The solid line represents the result from the multiple scattering calculation. The dashed line is the result from mutual boosting and shadowing. Data points were taken from \cite{Aamodt:2010cz}.}
\label{fig:dNchdeta1}
\end{figure}

\begin{figure}[ht] 
\centering
\includegraphics[width=0.95\linewidth]{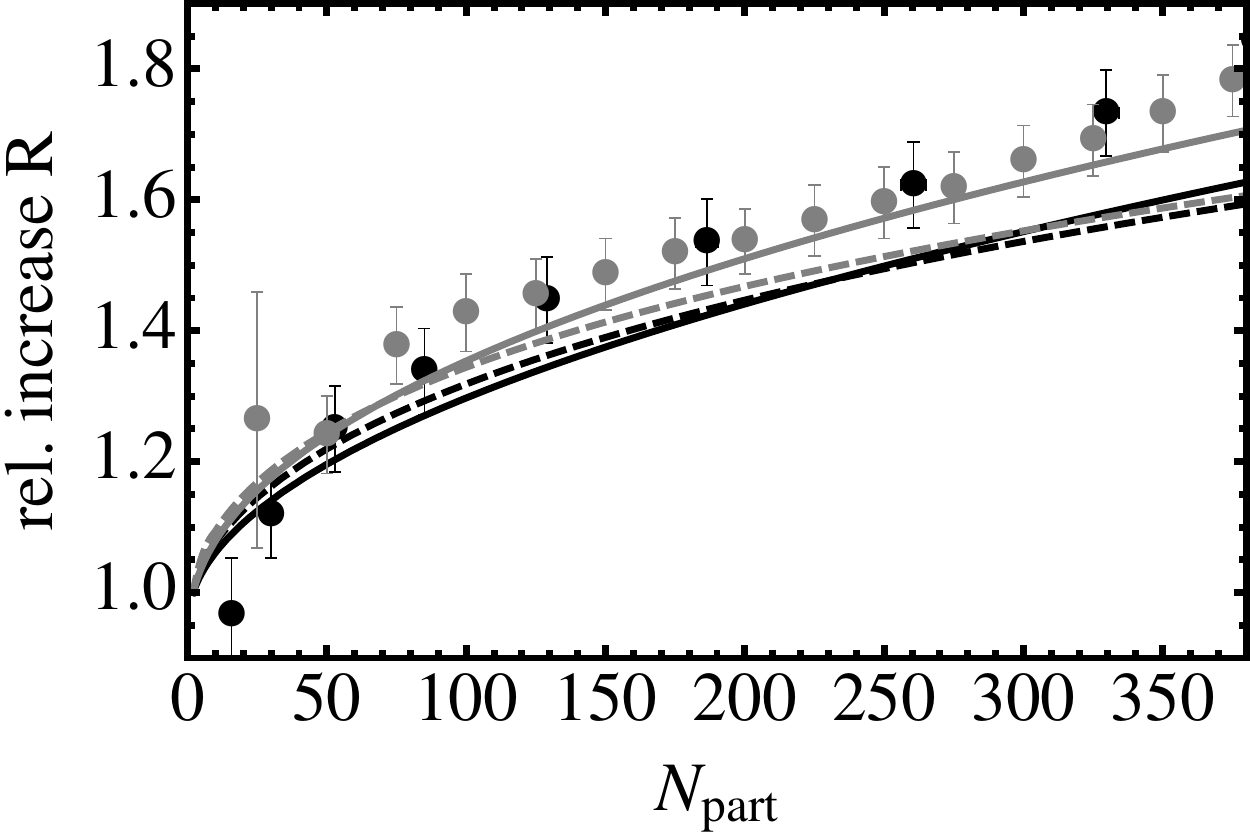}
\caption{Relative increase factor $R$ of the charged-particle multiplicity in AA collisions  with respect to the multiplicity in pp collisions as a function of $N_\mathrm{part}$ ($R = ((dN_\mathrm{ch}^{A+A}/d\eta)/(N_\mathrm{part}/2))$ / $(dN_\mathrm{ch}^{p+p}/d\eta)$). The universal increase of the data (gray=RHIC, black=LHC) is reasonably well reproduced by the corresponding calculations (solid lines: multiple scattering only, dashed lines: mutual boosting and shadowing).}
\label{fig:universal_increase}
\end{figure}

In Figs.~\ref{fig:dNchdeta2} and \ref{fig:dNchdeta1} we show the resulting pseudorapidity densities per ($N_\mathrm{part}/2$)  at $\eta=0$ as functions of the number of participants. Recall that the number of participants has been related to the average profile function. Figure~\ref{fig:universal_increase} shows that the universal, i.e., $\sqrt{s_{NN}}$ independent, relative increase of the measured charged-particle multiplicity $((dN_\mathrm{ch}^{A+A}/d\eta)/(N_\mathrm{part}/2))$ with centrality is reasonably well reproduced by the calculations. This universal feature is of geometrical origin. 

\section{Discussion of the results}
We found in Ref.~\cite{Pirner:2011ab} that pp and AA collisions can be described by light-cone distributions which are derived from the maximum-entropy principle and contain as parameters a mean transverse temperature and a softness. Our picture underlying this description is very simple: After an exchange of small transverse momenta the parton content of the proton is liberated, and parton-hadron duality relates the produced pions to this light-cone distribution.  One can generalize this picture to AA collisions letting the semi-inclusive cross sections scale with the number of participant nucleons; see Eq.~\ref{eq:npart_scaling}. However, this was not sufficient to explain the experimental data. In addition, it was necessary to use the increased mean transverse momentum observed experimentally in order to describe nuclear collisions. 
We have demonstrated in this paper that one can derive the parameters of the statistical model of nucleus-nucleus collisions from the parameters of pp collisions with the geometry and the dynamics of parton rescattering as input.

Let us review our assumptions. The derivation of the average path length of a parton in the target nucleus uses the number of participants as weight factor. This number depends weakly on the inelastic pp cross section and therefore induces small differences in the centrality dependences of the average profile functions of Au-Au and Pb-Pb collisions at RHIC and LHC energies; see~Fig.~1. Only a minor variation is observed in experiment. Perhaps here an improvement is possible. 

Cosmic ray physicists have speculated for a long time that the inelasticity parameter $K$ may strongly decrease with increasing cm energy, see, e.g., \cite{Wibig:2011he}.
We observe such a behavior rather clearly. Here a theoretical explanation is necessary. 

We consider the intrinsic momenta of the gluons to be described by the LCP distribution in order to average the dipole cross section and the gluon structure function. The LCP distribution comes from a statistical model for the nonequilibrium distribution of the gluons. It is an assumption that the intrinsic gluons have the same transverse momenta as the liberated gluons.

For the mutual boosting estimate the shadowing effect is taken into account very crudely. Furthermore, it is subtle that these two effects are counteracting each other leading to the same result as the crude calculation from parton multiple scattering. This is not expected at all energies and should be explored in more detail.
 
The simple features of the presented calculation are the most convincing: The multiplicity at pseudorapidity $\eta=0$ is roughly proportional to the product $L_{\bot}^2 \langle p_{\bot,\mathrm{tot}}^2\rangle$ \cite{Pirner:2011ab}.  Therefore, the geometrical dependence of transverse momentum broadening is clearly visible in the behavior of the rapidity distributions. The same dependence of the rapidity distributions on centrality in Au-Au at RHIC and Pb-Pb at the LHC confirms this finding.  There is a clear increase of the magnitude of the rapidity distribution at $\eta=0$ for the different energies. At the LHC ($\sqrt{s_{NN}} = \unit[2.76]{TeV}$) one obtains roughly a twice larger value for all centralities than at RHIC ($\sqrt{s_{NN}} =\unit[0.2]{TeV}$). This factor arises from the twice as large intrinsic mean $\langle p_{\bot}^2\rangle$ in pp collisions (see Table~I) and from the twice as large momentum broadening $\Delta  p_{\bot}^2$ from multiple scattering reflected in the energy dependence of the transport parameter $\langle \sigma p_{\bot}^2 \rangle$. 

So our presentation has been able to separate the geometrical from the dynamical effects. Very recently we received a preprint \cite{Schenke:2012hg} simulating rapidity distribution at $\eta=0$ including event-to-event fluctuations. On average, the above authors reproduce the same geometrical picture and saturation, including a microscopic picture of the initial gluon plasma fields.
 
\begin{acknowledgments} 
This work was supported by the Alliance Program of the Helmholtz Association, Contract No. HA216/EMMI "Extremes of Density and Temperature: Cosmic Matter in the Laboratory", by DFG (Germany) and CONICYT (Chile) Grant No. RE 3513/1-1, and by FONDECYT (Chile) Grant No. 1090291.  
\end{acknowledgments}
 
\appendix



\bibliography{universal11final}





\end{document}